\documentclass[twocolumn,english,prl]{revtex4-1}
\usepackage{mathptmx}

\usepackage[T1]{fontenc}
\usepackage[latin9]{inputenc}
\usepackage{bm}
\usepackage{amsmath}
\usepackage{amssymb}
\usepackage{graphicx}
\usepackage{babel}
\begin{document}
\title{Cooperative states and shift in resonant scattering of an atomic ensemble}
\author{Ting Hsu, Kuan-Ting Lin, and Guin-Dar Lin}
\affiliation{Center for Quantum Science and Engineering, Department of Physics,
National Taiwan University, Taipei 10617, Taiwan}
\begin{abstract}
We investigate the spectral shift in collective forward scattering
for a cold dense atomic cloud. The shift, sometimes called collective
Lamb shift, results from resonant dipole-dipole interaction mediated
by real and virtual photon exchange, forming many-body states displaying
various super- and subradiant spectral behavior. The scattering spectrum
reflects the overall radiative behavior from these states. However,
it also averages out the radiative details associated with a single
collective state, causing ambiguity in explaining the origin of the
spectral shift and raising controversy on its scaling property. We
employ a Monte-Carlo simulation to study how the collective states
are occupied and contribute to emission. We thus distinguish two kinds
of collective shift that follow different scaling laws. One results
from dominant occupation of the near-resonant collective states. This
shift is usually small and insensitive to the density or the number
of participating atoms. The other comes from large spatial correlation
of dipoles, associated with the states of higher degree of emission.
This corresponds to larger collective shift that is approximately
linearly dependent on the optical depth. Our analysis provides not
only a novel perspective for the spectral features in collective scattering,
but also a possible resolution to the controversy on the scaling property
that has been reported elsewhere because of different origins.
\end{abstract}
\maketitle
\noindent Light-ensemble interaction has been an important topic drawing
continuing attention for recent years thanks to its fundamental interest
in quantum many-body physics and practical applications in various
areas, such as atomic clocks and metrology \citep{Takamoto2005,Ludlow2015},
sensing and precision measurement \citep{Kitching2011}, quantum simulation
\citep{Bloch2012}, quantum interface, memory, and network \citep{Duan2001,Lukin2003,Kimble2008,Komar2014}.
In various proposals, schemes based on atomic ensembles are expected
to have enhanced coupling strength for more efficient manipulation
by increasing the number and/or density of atoms. As the system becomes
sufficiently dense, on the other hand, cooperative effects due to
atoms' dipole-dipole interaction start to emerge, including super-
and subradiance, directional emission, frequency shift, and distortion
of line shape \citep{Guerin2017a}. These effects may, for instance,
cause unwanted decoherence in quantum control and degrade the precision
of optical atomic clocks \citep{Chang2004}. Thus, how to understand
the cooperativity in atomic ensembles not only provides insightful
perspectives for many-body physics in the presence of nontrivial competing
interactions, but also helps develop quantum optical devices and applications
more accurately.

One intriguing cooperative phenomenon is the emergence of the collective
Lamb shift, the many-body version of the ordinary Lamb shift. The
ordinary Lamb shift accounts for the vacuum fluctuations that perturb
the electron's orbital in an atom and introduce an energy shift \citep{Welton1948}.
Understanding of this shift has opened a new subject now known as
quantum electrodynamics. To correctly calculate the shift, contributions
of all transition processes including virtual ones need to be properly
dealt with. Similar consideration applies to many-body cases, where
both the real and virtual processes of photon exchange mediate the
dipole-dipole interaction, resulting in cooperative decay and energy
shift of the collective states. Recently, such phenomena have attracted
extensive attention, and have been discussed in various contexts including
atomic clouds \citep{Jennewein2016,Araujo2016,Roof2016,Bromley2016},
nano-layer gases \citep{Keaveney2012,Peyrot2018}, ensembles of nuclei
\citep{Rohlsberger2010}, trapped ions \citep{Meir2014}, and artificial
atoms \citep{Loo2013,Wen2019,Lin2019}. Another perspective views
this shift as coupling between collective states, leading to Rabi-like
excitation transfer among a few atoms \citep{Barredo2015,Browaeys2016}.

A commonly-used technique to detect the collective shift is through
scattering experiments, where one measures the emission spectrum while
sweeping the probe frequency, and extracts the shift of the spectral
peak. Though it is valid to consider only the lowest excited manifold
for weak probing, as in many experiments, there are $N$ such singly-excited
states with $N$ the number of atoms in the ensemble, and these states
are shifted differently owing to competing dipole-dipole interaction.
Unfortunately, the scattering spectrum only reflects the overall effect
of superposed contributions from individual collective states. Some
unrevealed spin orders relevant to the spectrum might be averaged
out. Further, there have been controversies regarding the scaling
nature of the collective shift in different geometries such as slabs
and ellipsoids. Recent experiments have reported large collective
shift in elongated atomic samples \citep{Roof2016,Jennewein2016}
but negligible shift in pancake-like clouds \citep{Corman2017}. Changing
the anisotropy also introduces unusual shift scaling \citep{Bromley2016}.
For some specific geometries and densities, the lineshape becomes
asymmetric and even displays two peaks in the profile, making the
determination of the shift ambiguous.

\begin{figure}
\begin{centering}
\includegraphics[width=1\columnwidth]{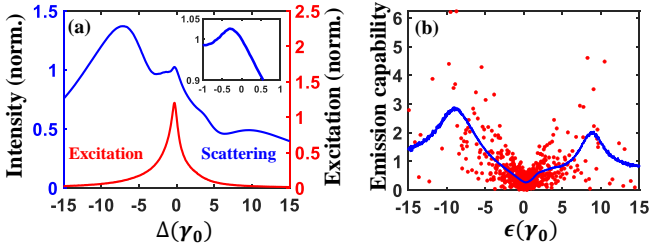}
\par\end{centering}
\caption{(a) Forward scattering (blue) and total excitation (red) spectrum
(ensemble averaged). Inset: zoomed plot near the central tip of the
scattering spectrum. The three peaks are located at $\Delta=-7.3\gamma_{0}$,
$-0.2\gamma_{0}$, and $+9.6\gamma_{0}$. (b) Emission capability
versus the eigenenergy $\epsilon$ (compatible to the state shift)
of the corresponding collective state. The scattered dots are from
a single configuration and the blue curve is from ensemble average.
The system used here is a cylindrical atomic ensemble of radius $R=2\pi c/\omega_{0}$
and length $L=3R$. The laser incident direction is along the axis
of the cylinder. The ensemble average is done by considering $>5000$
random configurations of atoms uniformly distributed inside the cylinder
with $\rho/k^{3}=0.3$, where $\rho$ is the number density. (The
scattering spectrum and excitation curves are normalized by the values
at $\Delta=0$. The emission capability is in arbitrary units.) \label{fig1:fsspect}}
\end{figure}

To better understand the underlying mechanisms, in this manuscript
we look into the roles of the many-body states by choosing an appropriate
orthogonal basis. We adopt a Monte-Carlo simulation by randomly distributing
atoms of controlled density within a given geometry, and use the standard
non-Hermitian Schrodinger equation approach in the weak-field (single-excitation)
limit. Note that this is a special case where the mathematical structure
is consistent with the classical dipole analysis. For each run, the
randomly picked spatial arrangement of atoms is called a configuration
in this manuscript. The calculated outcomes display huge fluctuations
from run to run, implying the sensitivity of the system to the actual
parameters. The robustness of the results lies in the ensemble averaged
profiles as we try many configurations. Let's first look at one of
the most illustrative examples corresponding to a uniform dense cylindrical
sample of radius comparable to a transition wavelength and length
over a few wavelengths. Fig.~\ref{fig1:fsspect} shows the corresponding
scattering spectrum, excitation distribution, and emission capability,
defined to characterize the spectral contribution of a state (see
following discussions). We can observe ``three peaks'' in the spectrum
by sweeping the probe detuning: the strongest left peak, the small
tip in the center, and the smooth hump on the right. Similar spectral
profiles have also been discussed in \citep{Sokolov2009,Sokolov2011}.
We find that the three peaks have different origins by taking into
consideration the roles played by the participating collective states.
Such diagnosis applies to general cases but the significance of the
three types of contribution may vary in different geometries and parameters.
For instance, the smooth hump on the blue side only emerges in cases
close to uniform distribution of atoms, and easily smears out in Gaussian
samples. The large shift on the left is more pronounced in elongated
systems, and gradually merges to the center peak as the anisotropy
reduces. Note that the central peak is also shifted, as shown in the
inset of Fig.~\ref{fig1:fsspect}(a). Most of the collective shift
measurements refer to either the left or the central peak (the right
hump is usually not very contrastive), which yields very different
scaling nature.

We consider an ensemble of $N$ two-level atoms randomly distributed
in a localized region of certain geometry with $\boldsymbol{r}_{i}$
denoting the location of the $i$th atom assumed to be fixed in space.
To probe the spectrum scattered by the singly-excited collective states
and the associated Lamb shift, these atoms are driven weakly by a
laser beam propagating along the $+z$ direction, where the detuning
$\Delta\equiv\omega-\omega_{0}\ll\omega_{0}$, wavevector $\bm{k}\equiv\omega\hat{z}/c\approx\omega_{0}\hat{z}/c$,
Rabi frequency $\Omega\ll\gamma_{0}$, with $\omega_{0}$ ($\omega$)
the atomic (laser) frequency and $\gamma_{0}$ the spontaneous emission
rate, and $c$ the speed of light. In the low-excitation limit, a
quantum state can be of the form $|\psi\rangle\approx|g_{1}\cdots g_{N}\rangle+\sum_{i}b_{i}|g_{1}\cdots e_{i}\cdots g_{N}\rangle$
with $|b_{i}|\ll1$, which satisfies $i\dot{\bm{B}}(t)=\mathbf{M}\bm{B}+\Omega\bm{D}$
\citep{Zhu2016}, where $\bm{B}=\left[b_{1},\ldots,b_{i},\ldots,b_{N}\right]^{\intercal},$
$\bm{D}=\left[e^{i\bm{k}\cdot\bm{r}_{1}},\ldots,e^{i\bm{k}\cdot\bm{r}_{i}},\ldots,e^{i\bm{k}\cdot\bm{r}_{N}}\right]^{\intercal}$,
and 
\begin{equation}
\bm{\mathbf{M}}=\begin{bmatrix}-\Delta-\frac{i\gamma_{0}}{2} & V_{12} & \cdots & V_{1N}\\
V_{21} & -\Delta-\frac{i\gamma_{0}}{2} & \cdots & \vdots\\
\vdots & \vdots & \ddots & \vdots\\
V_{N1} & \cdots & \cdots & -\Delta-\frac{i\gamma_{0}}{2}
\end{bmatrix}.\label{eq:matrixM}
\end{equation}
The dipole-dipole interaction is given by
\begin{multline}
V_{ij}=\frac{3\gamma_{0}}{4}\Big[-\left(1-\cos^{2}\theta_{ij}\right)\frac{e^{ikr_{ij}}}{kr_{ij}}+\\
\left(1-3\cos^{2}\theta_{ij}\right)\left(\frac{-ie^{ikr_{ij}}}{(kr_{ij})^{2}}+\frac{e^{ikr_{ij}}}{(kr_{ij})^{3}}\right)\Big],\label{eq:ddi}
\end{multline}
depending on the separation $r_{ij}=|\bm{r}_{j}-\bm{r}_{i}|$ between
two atoms $i$ and $j$, and the angle $\theta_{ij}$ between the
distance vector $\bm{r}_{j}-\bm{r}_{i}$ and the dipole orientation
assumed to be the $x$ direction for linear polarized driving field.
The dipole-dipole interaction couples the $N$ singly excited states,
causing various degrees of shift depending on the configuration of
distribution of the atoms. When $N$ is large, it is however unrealistic
and meaningless to measure the shift of each collective state. Nevertheless,
the scattering spectrum still catch the overall contributions, and
can help identify the emergence and order of magnitude of the collective
Lamb shift. But we still need to look into what are actually probed
for further diagnosis.

By solving $\bm{B}=-\Omega\mathbf{M}^{-1}\bm{D}$, we obtain the steady-state
solution $b_{i}$, which determines the intensity of the forward scattering
by $I\propto\left|\sum_{i}b_{i}e^{-i\bm{k}\cdot\bm{r}_{i}}\right|^{2}=\sum_{i}\left|b_{i}\right|^{2}+\sum_{i\neq j}b_{i}b_{j}^{*}e^{-i\bm{k}\cdot(\bm{r}_{i}-\bm{r}_{j})}$.
We identify two parts: the incoherent scattering term $\sum_{i}\left|b_{i}\right|^{2}$
and the coherent one $\sum_{i\neq j}b_{i}b_{j}^{*}e^{-i\bm{k}\cdot(\bm{r}_{i}-\bm{r}_{j})}$.
The former exactly corresponds to the total excitation of individual
atoms for a given probe detuning. The latter accounts for the spin-spin
correlation modulated by the spatial phases. The resultant spectral
profile of emission fluctuates drastically for a given spatial configuration
of atoms. We thus take the ensemble average over many configurations,
and generate a convergent coarse-grained profile, as shown in Fig.\,\ref{fig1:fsspect}.

To understand the spectrum, it is natural to relate the radiative
properties to the many-body states that are relevant. Note that the
coupling matrix $\mathbf{M}$ is symmetric but complex. In some previous
literature, it has been directly diagonalized to find the dynamics
for the system \citep{Li2013,Schilder2016,Guerin2017}. The collective
basis states thus found are however not orthogonal. Though the calculated
dynamics is correct, it fails to associate the radiative behavior
with specific states because they are seriously overlapped \citep{Li2013}.
Here, by noting that $\mathbf{M}=-\Delta\mathbf{I}+\mathbf{M}_{R}+i\mathbf{M}_{I}$,
where $\mathbf{I}$ is an $N\times N$ identity matrix, $\mathbf{M}_{R}$
and $\mathbf{M}_{I}$ are the real and imaginary parts of $\mathbf{M}$,
respectively, we choose the orthogonal eigenbasis that diagonalizes
only the real part $\mathbf{M}_{R}$ such that $\mathbf{M}_{R}^{D}=\mathbf{R}^{\intercal}\mathbf{M}_{R}\mathbf{R}$
with $\left[\mathbf{M}_{R}^{D}\right]_{ij}=\epsilon_{j}\delta_{ij}$.
The eigenvalue $\epsilon_{j}$ is associated with an eigenvector $\bm{R}_{j}$
(the $j$th column of $\mathbf{R}$), and can be understood as the
shift of the $j$th collective state $|\bm{R}_{j}\rangle=\sum_{s}\bm{\mathbf{R}}_{sj}|g_{1}\cdots e_{s}\cdots g_{N}\rangle$.

\begin{figure}
\begin{centering}
\includegraphics[width=1\columnwidth]{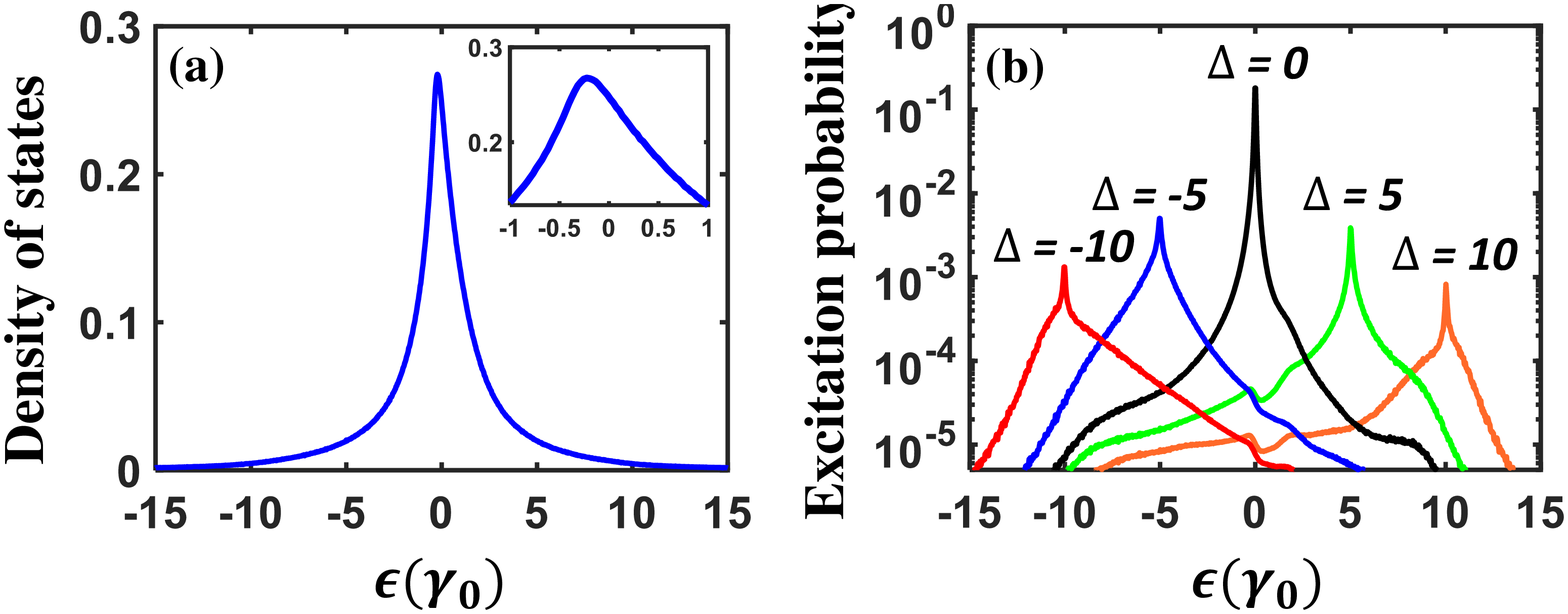}
\par\end{centering}
\caption{(a) Density of states (normalized by the area underneath). Inset:
zoomed plot around the tip, which is red shifted by about $0.2\gamma_{0}$
to the single atom's resonance. (b) Histogram of the excitation distribution
(arbitrary unit) of at various detunings. (These curves are also ensemble
averaged.) \label{fig2:excit}}
\end{figure}

Figure~\ref{fig2:excit}(a) shows the density of states as a function
of a state's shift $\epsilon$. The profile suggests that the majority
of the collective states ($\sim82.3\%$) have shift less than $5\gamma_{0}$.
The density-of-state profile has a peak slightly shifted to the red
side of the resonant one ($\epsilon=0$), which results in the shift
of the central peak. We can now discuss the excitation $\left|p_{j}\right|^{2}$
of the state $|\bm{R}_{j}\rangle$ by applying a pumping laser of
frequency $\omega=\omega_{0}+\Delta$. The steady-state solution $\bm{B}=\sum_{j}p_{j}\bm{R}_{j}$,
and therefore $p_{j}=\left(\mathbf{R}^{-1}\bm{B}\right)_{j}$. We
plot the excitation histogram of different shift for various detuning
$\Delta$ in Fig.~\ref{fig2:excit}(b). As expected, the probe field
excites only those states resonant to the laser frequency ($\Delta\approx\epsilon$)
with a linewidth comparable to $\gamma_{0}$. The peak values of excitation
also drops significantly for large $|\Delta|$ since the density of
states is small of large $|\epsilon|$. We emphasize that the excitation
behavior shown in Fig.~\ref{fig2:excit}(b) is only valid in this
choice of orthogonal basis (from $\mathbf{M}_{R}$).

As we sweep the detuning, we might have expected that there would
be a significant peak in the center of the spectrum, corresponding
to the most probable excited states indicated by Fig.~\ref{fig2:excit}(a).
In fact, this is not the case here. Those most-populated states only
contribute to a small tip in the spectrum. By contrast, the most significant
peak appears on the red side ($\Delta\approx-7.3\gamma_{0}$ in Fig.~\ref{fig1:fsspect}(a)).
But the corresponding peak excitation is lower by two orders of magnitude
than that of $\Delta=0$. This suggests that these states, though
just a few of them, are much more radiative compared to the majority
of the collective states. To quantify for the spectral contribution
of a state, we define for the state $|\bm{R}_{j}\rangle$ the emission
capability 
\begin{align}
\Pi_{j} & =\left|\sum_{i}\bm{R}_{j}^{(i)}e^{-i\bm{k}\cdot\bm{r}_{i}}\right|,\label{eq:emcap}
\end{align}
where $\bm{R}_{j}^{(i)}$ is the $i$th element (corresponding to
the $i$th atom) of the eigenvector $\bm{R}_{j}$. Since $I\propto\left|\sum_{i,j}p_{j}\bm{R}_{j}^{(i)}e^{-i\bm{k}\cdot\bm{r}_{i}}\right|^{2}$,
if $\Pi_{j}$ is large (small), it can be expected that the state
$|\bm{R}_{j}\rangle$ takes part significantly (insignificantly) in
the scattering spectrum. It helps us to diagnose the behavior of each
collective state and provides a novel perspective regarding super-
and subradiance without explicitly taking into account the collective
decay rates.

In Fig.~\ref{fig1:fsspect}(b), we plot the emission capability for
a single configuration (red dots) and on average (blue line). We find
that the emission capacity curve has two peaks around $\epsilon\approx-8.2\gamma_{0}$
and $\epsilon\approx+9.1\gamma_{0}$, implying that these states are
highly capable of emission. The left peak does give rise to the most
visible peak in the spectrum of Fig.~\ref{fig1:fsspect}(a). The
right peak is also responsible for a small hump at the corresponding
detuning of the spectral profile. This hump on the blue side is however
less evident because the corresponding states are less populated.
On the other hand, the states of small shift have low emission capability
even though they are mostly populated. This ``M-shaped'' feature
of emission capability is generally observed in all atomic configurations,
even in various geometries and densities. But the actual spectral
curves are still determined by considering overall the emission capability,
density of states, excitation, and cross-term interference of the
collective states in detail, which vary from sample to sample of different
parameters.

\begin{figure}
\begin{centering}
\includegraphics[width=1\columnwidth]{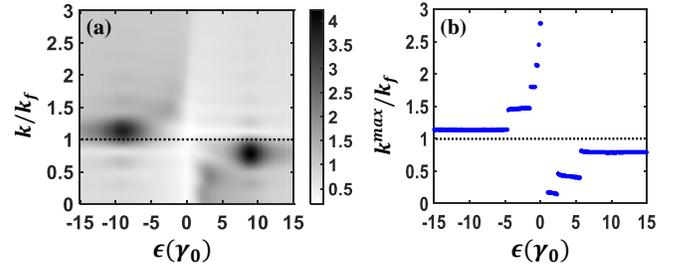}
\par\end{centering}
\caption{(a) Fourier spectrum of the spatial frequency versus eigenenergy with
the color darkness (arbitrary unit) represents $F(\epsilon,k_{f})$.
(b) Maximal spatial frequency extracted from (a).\label{fig3:order}}
\end{figure}

Note that the probe detuning has nothing to do with the structure
of the collective states $|\bm{R}_{j}\rangle$, which are determined
solely by the interaction detail, $\mathbf{M}_{R}$. But the probe
sets a detuning window that selects which collective states are pumped.
Therefore, the scattered signal must reflect the spatial order of
the picked states. The definition of emission capability Eq.~(\ref{eq:emcap})
is a reminiscence of the Fourier transform of $\bm{R}_{j}$. We further
define 
\begin{equation}
\tilde{F}_{j}(k_{f})=\sum_{i}\bm{R}_{j}^{(i)}e^{-ik_{f}z_{i}},\label{eq:ft}
\end{equation}
which bares the information of components with ``spatial frequencies''
$k_{f}$ for a given state $|\bm{R}_{j}\rangle$. Here, we only focus
on the spatial variation along the $+z$ propagation axis for forward
scattering. We show the ensemble-averaged Fourier spectrum $F(\epsilon,k_{f})\equiv\left\langle \left|\tilde{F}_{j}(k_{f})\right|\right\rangle $
(averaged over $j$ of similar $\epsilon$) in Fig.~\ref{fig3:order}(a).
The dotted line refers to $k_{f}=k=2\pi/\lambda$, taken as a reference
of spatial order in periods of $\lambda$. Further, the pattern of
$F(\epsilon,k_{f})$ on both the red and blue sides appears to be
stripe-like. For a given $\epsilon$, we identify the spatial frequency
$k_{f}^{max}(\epsilon)$ that maximizes $F(\epsilon,k_{f})$, plotted
in Fig.~\ref{fig3:order}(b). We observe that the states who possess
the most distinct spatial ordering (darkest spots in Fig.~\ref{fig3:order}(a))
coincide with those of highest emission capability. This can be expected
because the characteristic $k_{f}^{max}(\epsilon)$ of these states
is closest to $k$, i.e., matching the spatial modulation of the incident
light. This spatial ordering implies the cooperative nature of superradiance,
and thus contributes significantly to the emission. By contrast, the
states of small $|\epsilon|$ do not display a clear spatial order.
This is due to random phase cancellation from averaging out a huge
amount of such collective states. Note that the discreteness of $k_{f}^{max}$
is a reminiscence of the standing wave conditions, for which we find
the spatial frequency gap $\Delta k_{f}^{max}\sim2\pi/L$. But we
do not have exactly $k_{f}^{max}=k$. This might be due to the size
effect of a finite cylinder.

\begin{figure}
\begin{centering}
\includegraphics[width=1\columnwidth]{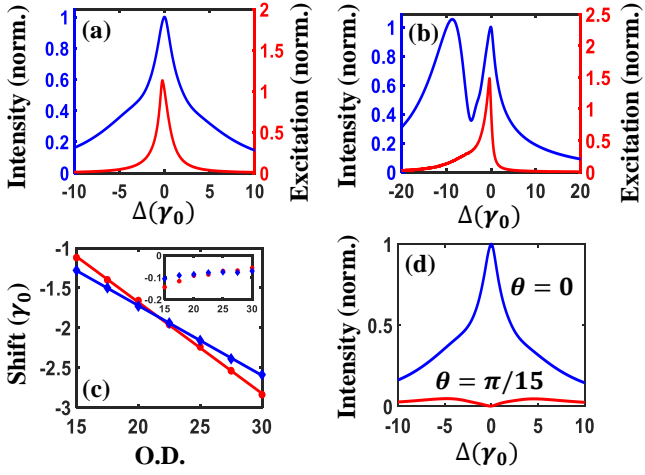}
\par\end{centering}
\caption{(a) Spherical Gaussian ensemble with $N=1000$ and $\sigma_{x,y,z}=\lambda_{0}$.
(b) Cigar-shaped ensemble with $N=1000$, $\sigma_{x}=\sigma_{y}=0.5\lambda_{0}$
and $\sigma_{z}=4\lambda_{0}$. (c) Collective shift for varied OD.
Red solid circles correspond to changing the system size by fixing
the aspect ratio $\sigma_{x}:\sigma_{y}:\sigma_{z}=1:1:10$ and $\rho_{0}/k^{3}=0.01$.
Blue solid diamonds correspond to changing the density by fixing $\sigma_{x}=\sigma_{y}=1.5\lambda_{0}$,
and $\sigma_{z}=15\lambda_{0}$. Inset: collective shift of the central
spectral peak. (d) Scattering spectrum at $\theta=\frac{\pi}{15}$
reveals the two largely-shifted peaks as $\theta=0$ shows only one
peak. The parameters used are identical to (a). {[}All the spectral
and excitation curves are normalized by the values at $\Delta=0$
(and $\theta=0$ in (d)).{]}\label{fig4:scaling}}
\end{figure}

Now we discuss the collective shift in the cases of different geometry
and density. We focus on two exemplary cases: spherical and cigar-shaped
samples. To better approximate the actual ensembles in experiments,
we consider atomic clouds of Gaussian density distribution: $\rho(\bm{r})=\rho_{0}\exp[-(x^{2}/\sigma_{x}^{2}+y^{2}/\sigma_{y}^{2}+z^{2}/\sigma_{z}^{2})/2]$,
where $\rho_{0}=(2\pi)^{3/2}(\sigma_{x}\sigma_{y}\sigma_{z})^{-1}N$
is the peak density. For spherical samples, in Fig.~\ref{fig4:scaling}(a)
we present the scattering spectrum and total excitation profile of
a typical case with a peak density $\rho_{0}/k^{3}=0.26$, $\sigma_{x,y,z}=\lambda_{0}\equiv2\pi c/\omega_{0}$,
and $N=1000$. Here, we only observe the central peak, apparently
due to significant population of the near-resonant state. The one-peak
feature remains for spherical Gaussian samples of $\sigma_{x,y,z}$
over a few wavelengths (not shown). In these cases, the occupation
of the most radiative states is very low.

The left peak becomes more evident only when the radiative spatial
order can be supported by the medium. We thus expect to observe a
clear signal in elongated samples. We then consider a cigar-shaped
sample with $\sigma_{x}=\sigma_{y}=0.5\lambda_{0}$ and $\sigma_{z}=4\lambda$
while keeping the same number and peak density. Fig.~\ref{fig4:scaling}(b)
shows the spectrum presenting a significant left peak around $\Delta=-9.3\gamma_{0}$
together with the central one. By checking the excitation curve, we
find strong asymmetry with respect to $\Delta=0$. The near-resonant
states are still highly populated. But the excitation of red-shifted
states is considerably larger than that in the spherical case. That
of the blue-shifted states is now strongly suppressed. This explains
the two-peak spectral profile.

We further examine the dependence of the collective shift on the density
in terms of the optical depth $\text{OD}=3N/(2k^{2}\sigma_{x}\sigma_{y})$
in the cigar-shaped cases of $\sigma_{x}:\sigma_{y}:\sigma_{z}=1:1:10$.
We vary OD in two ways while keeping the aspect ratio the same: One
is to fix the peak density and change the size. The other is to fix
the size and change the density. The results are plotted in Fig.~\ref{fig4:scaling}(c).
Both of them show approximately linear relations versus OD but not
exactly coincide. Their difference seems to reflect the finite-size
effect, which needs further investigation. We also demonstrate the
central-peak shift, usually smaller than $\gamma_{0}$ by an order
of magnitude, showing very different dependence on OD. Note that this
shift originates mainly from the central peak of the density of states,
which can be obtained by looking at the histogram the eigenenergy
distribution. Since the matrix $\mathbf{M}_{R}$ is traceless, these
eigenenergies must add up to be zero, thus restricting the shift from
varying sensitively against OD, and presenting no linear scaling.

Finally, we discuss the interesting angular dependence of the scattering
spectrum as shown in Fig.~\ref{fig4:scaling}(d) and reported in
\citep{Zhu2016}. At $\Delta=0$, the scattering intensity is mostly
owing to near-resonant states, consisting of both incoherent and coherent
contributions. It is usually the coherent part that dominates, causing
of the forward directional emission with enhanced intensity. However,
since these states ($\epsilon\approx0$) does not possess distinct
spatial order, the coherent scattering is only confined within a narrow
angle about the forward direction. When the spectrum is detected at
a finite but small angle, the central peak drops rapidly, revealing
the two largely-shifted peaks. Since these two peaks correspond to
states presenting distinct spatial ordering, and hence are more robust
against the detection angle.

To sum up, we have investigated the collective shift by Monte-Carlo
simulation, and studied the state properties in the ensemble-average
manner. By examining the spectral contributions associated with the
collective states, which are strongly related to the spatial modulation
of the spin orders, we have identified two types of collective shift
presenting different scaling laws. We believe that our work can provide
an insightful perspective for recent experiments. As future outlook,
we will investigate the roles of many-body states in terms of spatial
orders on the collective decay and linewidth. Also, we find that the
relaxation dynamics of timed Dicke states shares similar mathematical
structures with our approach \citep{Li2013}. We will also look into
the connection to the single-photon relaxation experiments.
\begin{acknowledgments}
We thank the support from MOST of Taiwan under Grant No.~109-2112-M-002-022
and National Taiwan University under Grant No.~NTU-CC-109L892006.
GDL thanks Ying-Cheng Chen, Hsiang-Hua Jen, and Ming-Shien Chang for
valuable discussion and feedback.
\end{acknowledgments}

\bibliographystyle{apsrev4-1}
\bibliography{citation}

\end{document}